\documentclass[twocolumn,A4]{article}
\usepackage[dvips]{graphics}
\usepackage{setspace}
\usepackage{color}
\definecolor{gold}{rgb}{0.85,0.66,0}
\definecolor{dblue}{rgb}{0,0,0.8}
\begin{document}
\onecolumn
\begin{center}
{\bf{\Large {\textcolor{gold}{XOR gate response in a mesoscopic ring 
with embedded quantum dots}}}}\\
~\\
{\textcolor{dblue}{Santanu K. Maiti}}$^{1,2,*}$ \\
~\\
{\em $^1$Theoretical Condensed Matter Physics Division,
Saha Institute of Nuclear Physics, \\
1/AF, Bidhannagar, Kolkata-700 064, India \\
$^2$Department of Physics, Narasinha Dutt College,
129, Belilious Road, Howrah-711 101, India} \\
~\\
{\bf Abstract}
\end{center}
We address XOR gate response in a mesoscopic ring threaded by a magnetic
flux $\phi$. The ring, composed of identical quantum dots, is symmetrically 
attached to two semi-infinite one-dimensional metallic electrodes and two 
gate voltages, viz, $V_a$ and $V_b$, are applied, respectively, in each arm 
of the ring which are treated as the two inputs of the XOR gate. The 
calculations are based on the tight-binding model and the Green's function 
method, which numerically compute the conductance-energy and current-voltage 
characteristics as functions of the ring-electrodes coupling strengths, 
magnetic flux and gate voltages. Quite interestingly it is observed that, 
for $\phi=\phi_0/2$ ($\phi_0=ch/e$, the elementary flux-quantum) a high 
output current ($1$) (in the logical sense) appears if one, and only one, 
of the inputs to the gate is high ($1$), while if both inputs are low 
($0$) or both are high ($1$), a low output current ($0$) appears. It 
clearly demonstrates the XOR behavior and this aspect may be utilized 
in designing the electronic logic gate. 

\vskip 1cm
\begin{flushleft}
\end{flushleft}
\vskip 4.1in
\noindent
{\bf ~$^*$Corresponding Author}: Santanu K. Maiti

Electronic mail: santanu.maiti@saha.ac.in

\newpage
\twocolumn

\section{Introduction}

In the present age of nanoscience and technology quantum confined model
systems are used extensively in electronic as well as spintronic engineering 
since these simple looking systems are the fundamental building blocks of 
designing nano devices. A mesoscopic normal metal ring is one such promising 
example of quantum confined systems. Here we will explore the electron 
transport through a mesoscopic ring, composed of identical quantum dots and
attached to two external electrodes, the so-called electrode-ring-electrode 
bridge, and show how such a simple geometric model can be used to design 
a logic gate. The theoretical description of electron transport in a 
bridge system has been followed based 
on the pioneering work of Aviram and Ratner~\cite{aviram}. Later, many 
excellent experiments~\cite{tali,reed1,reed2} have been done in several 
bridge systems to understand the basic mechanisms underlying the electron 
transport. Though in literature many theoretical~\cite{orella1,orella2,
nitzan1,nitzan2,new,muj1,muj2,walc2,walc3,cui} as well as experimental 
papers~\cite{tali,reed1,reed2} on electron transport are available, yet 
lot of controversies are still present between the theory and experiment, 
and the complete knowledge of the conduction mechanism in this scale is 
not very well established even today. 
The electronic transport in the ring significantly depends on the 
ring-to-electrodes interface structure. By changing the geometry one can
tune the transmission probability of an electron across the ring. This
is solely due to the quantum interference effect among the electronic 
waves traversing through different arms of the ring. Furthermore, the electron 
transport through the ring can be modulated in other way by tuning the 
magnetic flux, the so-called Aharonov-Bohm (AB) flux, that threads the ring. 
The AB flux threading the ring can change the phases of the wave functions 
propagating along the different arms of the ring leading to constructive or 
destructive interferences, and accordingly the transmission amplitude 
changes~\cite{baer2,baer3,tagami,walc1,baer1}. Beside these factors, 
ring-to-electrodes coupling is another important issue that controls the 
electron transport in a meaningful way~\cite{baer1}. All these are the 
key factors which regulate the electron transmission in the 
electrode-ring-electrode bridge system and these effects have to be taken
into account properly to reveal the transport mechanisms. 

The aim of the present paper is to describe the XOR gate response in a 
mesoscopic ring threaded by a magnetic flux $\phi$. The ring is contacted 
symmetrically to the electrodes, and the two arms of the ring are subjected 
to two gate voltages $V_a$ and $V_b$, respectively (see Fig.~\ref{xor}) 
those are treated as the two inputs of the XOR gate. Here we adopt a simple 
tight-binding model to describe the system and all the calculations are 
performed numerically. We address the XOR behavior by studying the 
conductance-energy and current-voltage characteristics as functions of the 
ring-electrodes coupling strengths, magnetic flux and gate voltages. Our
study reveals that for a particular value of the magnetic flux, 
$\phi=\phi_0/2$, a high output current ($1$) (in the logical sense) is 
available if one, and only one, of the inputs to the gate is high ($1$),
while if both the inputs are low ($0$) or both are high ($1$), a low output
current ($0$) is available. This phenomenon clearly demonstrates the XOR
behavior which may be utilized in manufacturing the electronic logic gate.
To the best of our knowledge the XOR gate response in such a simple system
has not been described earlier in the literature.

The paper is organized as follow. Following the introduction 
(Section $1$), in Section $2$, we present the model and the theoretical 
formulations for our calculations. Section $3$ discusses the significant 
results, and finally, we summarize our results in Section $4$.

\section{Model and the synopsis of the theoretical background}

\begin{figure}[ht]
{\centering \resizebox*{7.7cm}{6.5cm}{\includegraphics{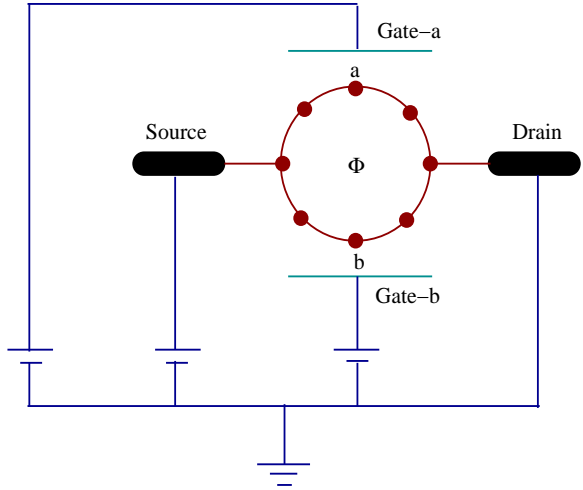}}\par}
\caption{(Color online). The scheme of connections with the batteries 
for the operation of the XOR gate. A mesoscopic ring with embedded quantum 
dots (filled red circles) is attached to two semi-infinite one-dimensional 
metallic electrodes, viz, source and drain. The gate voltages $V_a$ and 
$V_b$, those are variable, are applied in the dots $a$ and $b$ of the 
two arms, respectively. The source and the gate voltages are applied
with respect to the drain.}
\label{xor}
\end{figure}

We begin by referring to Fig.~\ref{xor}. A mesoscopic ring, composed 
of identical quantum dots (filled red circles) and threaded by a 
magnetic flux $\phi$, is attached symmetrically to two semi-infinite 
one-dimensional metallic electrodes. The ring is placed between two 
gate electrodes, viz, gate-a and gate-b. These gate electrodes are ideally
isolated from the ring and can be regarded as two parallel plates of
a capacitor. In our present scheme we assume that the gate voltages
each operate on the dots nearest to the plates only. While, in 
complicated geometric models, the effect must be taken into account 
for the other dots, though the effect becomes too small. The dots $a$ 
and $b$ in the two arms of the ring are subjected to the gate voltages 
$V_a$ and $V_b$, respectively, and these are treated as the two inputs 
of the XOR gate. The actual scheme of connections with the batteries
for the operation of the XOR gate is clearly presented in the
figure (Fig.~\ref{xor}), where the source and the gate voltages are 
applied with respect to the drain.

Based on the Landauer conductance formula~\cite{datta,marc} we determine the
conductance ($g$) of the ring. At very low temperature and bias voltage it 
can be expressed in the form,
\begin{equation}
g=\frac{2e^2}{h} T
\label{equ1}
\end{equation}
where $T$ gives the transmission probability of an electron through the ring. 
This $(T)$ can be represented in terms of the Green's function of the 
ring and its coupling to the two electrodes by the 
relation~\cite{datta,marc},
\begin{equation}
T={\mbox{Tr}} \left[\Gamma_S G_{R}^r \Gamma_D G_{R}^a\right]
\label{equ2}
\end{equation}
where $G_{R}^r$ and $G_{R}^a$ are respectively the retarded and advanced
Green's functions of the ring including the effects of the electrodes.
The parameters $\Gamma_S$ and $\Gamma_D$ describe the coupling of the
ring to the source and drain, respectively. For the full system i.e., 
the ring, source and drain, the Green's function is defined as,
\begin{equation}
G=\left(E-H\right)^{-1}
\label{equ3}
\end{equation}
where $E$ is the injecting energy of the source electron. To Evaluate
this Green's function, the inversion of an infinite matrix is needed since
the full system consists of the finite ring and the two semi-infinite 
electrodes. However, the entire system can be partitioned into sub-matrices 
corresponding to the individual sub-systems and the Green's function for 
the ring can be effectively written as,
\begin{equation}
G_{R}=\left(E-H_{R}-\Sigma_S-\Sigma_D\right)^{-1}
\label{equ4}
\end{equation}
where $H_{R}$ is the Hamiltonian of the ring that can be expressed within 
the non-interacting picture like,
\begin{eqnarray}
H_{R} & = & \sum_i \left(\epsilon_{i0} + V_a \delta_{ia} + V_b \delta_{ib} 
\right) c_i^{\dagger} c_i \nonumber \\
 & + & \sum_{<ij>} t \left(c_i^{\dagger} c_j e^{i\theta}+ c_j^{\dagger} 
c_i e^{-i\theta}\right)
\label{equ5}
\end{eqnarray}
In this Hamiltonian $\epsilon_{i0}$'s are the on-site energies for all the 
sites (filled red circles) $i$, except the sites $i=a$ and $b$ where the 
gate voltages $V_a$ 
and $V_b$ are applied, those are variable. These gate voltages can be 
incorporated through the site energies as expressed in the above 
Hamiltonian. $c_i^{\dagger}$ ($c_i$) is the creation (annihilation) 
operator of an electron at the site $i$ and $t$ is the nearest-neighbor 
hopping integral. $\theta=2 \pi \phi/N \phi_0$ is the phase factor due 
to the flux $\phi$, where $N$ represents the total number of sites/dots 
in the ring. Similar kind of tight-binding Hamiltonian is also used, except 
the phase factor $\theta$, to describe the semi-infinite one-dimensional 
perfect electrodes where the Hamiltonian is parametrized by constant on-site 
potential $\epsilon_0$ and nearest-neighbor hopping integral $t_0$. The ring 
is coupled to the electrodes by the parameters $\tau_S$ and $\tau_D$, where 
they (coupling parameters) correspond to the coupling strengths with the 
source and drain, respectively. The parameters $\Sigma_S$ and $\Sigma_D$ in 
Eq.~(\ref{equ4}) represent the self-energies due to the coupling of the ring 
to the source and drain, respectively, where all the informations of this 
coupling are included into these self-energies.

To evaluate the current ($I$), passing through the ring, as a function 
of the applied bias voltage ($V$) we use the relation~\cite{datta},
\begin{equation}
I(V)=\frac{e}{\pi \hbar}\int \limits_{E_F-eV/2}^{E_F+eV/2} T(E)~ dE
\label{equ8}
\end{equation}
where $E_F$ is the equilibrium Fermi energy. Here we make a realistic
assumption that the entire voltage is dropped across the ring-electrode
interfaces, and it is examined that under such an assumption the $I$-$V$
characteristics do not change their qualitative features. 

In this presentation, all the results are computed only at absolute zero 
temperature. These results are also valid even for some finite (low) 
temperatures, since the broadening of the energy levels of the ring due 
to its coupling with the electrodes becomes much larger than that of the 
thermal broadening~\cite{datta}. On the other hand, at high temperature 
limit, all these phenomena completely disappear. This is due to the fact 
that the phase coherence length decreases significantly with the rise of 
temperature where the contribution comes mainly from the scattering on
phonons, and accordingly, the quantum interference effect vanishes. 
For the sake of simplicity, we take the unit $c=e=h=1$ in our present 
calculations. 

\section{Results and discussion}

To illustrate the results, let us first mention the values of the different
parameters used for the numerical calculations. In the ring, the on-site
energy $\epsilon_{i0}$ is taken as $0$ for all the sites $i$, except the
sites $i=a$ and $b$ where the site energies are taken as $V_a$ and $V_b$, 
\begin{figure}[ht]
{\centering \resizebox*{8cm}{7cm}{\includegraphics{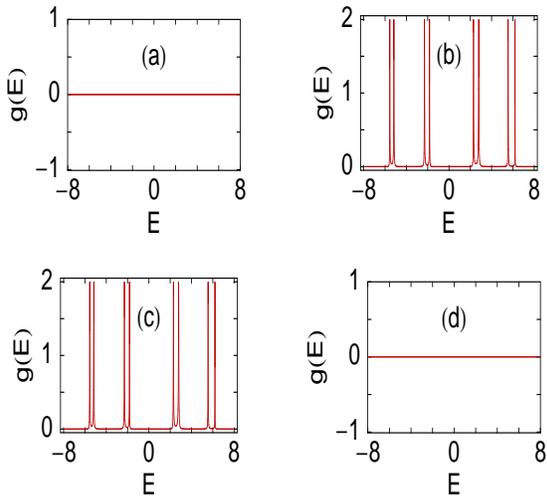}}\par}
\caption{(Color online). Conductance $g$ as a function of the energy $E$ 
for a mesoscopic ring with $N=8$ and $\phi=0.5$ in the limit of 
weak-coupling. (a) $V_a=V_b=0$, (b) $V_a=2$ and $V_b=0$, (c) $V_a=0$ 
and $V_b=2$ and (d) $V_a=V_b=2$.}
\label{condlow}
\end{figure}
respectively, and the nearest-neighbor hopping strength $t$ is set
to $3$. On the other hand, for the side attached electrodes the 
on-site energy ($\epsilon_0$) and the nearest-neighbor hopping strength 
($t_0$) are fixed to $0$ and $4$, respectively. The Fermi energy $E_F$ is
set to $0$. To narrate the coupling effect, throughout the study we 
focus our results for the two limiting cases depending on the strength 
of the coupling of the ring to the source and drain. Case $I$: The 
weak-coupling limit. It is described by the condition $\tau_{S(D)} << t$. 
For this regime we choose $\tau_S=\tau_D=0.5$. Case $II$: The 
strong-coupling limit. This is specified by the condition $\tau_{S(D)} 
\sim t$. In this particular regime, we set the values of the parameters 
as $\tau_S=\tau_D=2.5$. The key controlling parameter for all these 
calculations is the magnetic flux $\phi$ which is set to $\phi_0/2$ i.e.,
$0.5$ in our chosen unit $c=e=h=1$.

In Fig.~\ref{condlow} we present the conductance-energy ($g$-$E$) 
characteristics for a mesoscopic ring with $N=8$ in the limit of 
weak-coupling, where (a), (b), (c) and (d) correspond to the results
for the different gate voltages. When both the two inputs $V_a$ and $V_b$ 
are identical to zero i.e., both the inputs are low, the conductance $g$ 
\begin{figure}[ht]
{\centering \resizebox*{8cm}{7cm}{\includegraphics{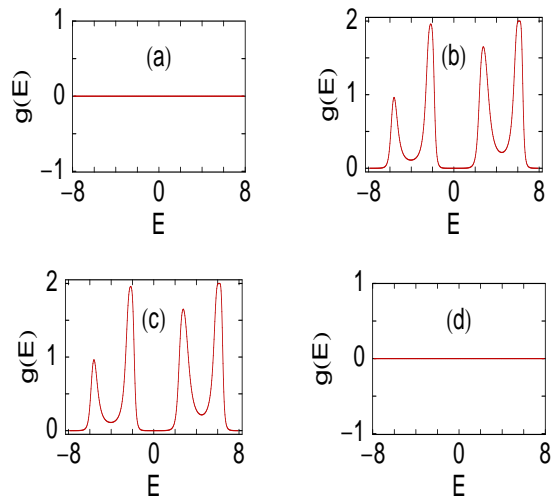}}\par}
\caption{(Color online). Conductance $g$ as a function of the energy $E$ 
for a mesoscopic ring with $N=8$ and $\phi=0.5$ in the limit of 
strong-coupling. (a) $V_a=V_b=0$, (b) $V_a=2$ and $V_b=0$, (c) $V_a=0$ 
and $V_b=2$ and (d) $V_a=V_b=2$.}
\label{condhigh}
\end{figure}
becomes exactly zero (Fig.~\ref{condlow}(a)) for all energies. This reveals 
that the electron cannot conduct through the ring. Similar response is also 
observed when both the two inputs are high i.e., $V_a=V_b=2$, and in this 
case also the ring does not allow to pass an electron from the source to
the drain (Fig.~\ref{condlow}(d)). On the other hand, for the cases 
where any one of the two inputs is high and other is low i.e., either 
$V_a=2$ and $V_b=0$ (Fig.~\ref{condlow}(b)) or $V_a=0$ and $V_b=2$
(Fig.~\ref{condlow}(c)), the conductance exhibits fine resonant peaks 
for some particular energies. Thus for both these two cases the electron 
conduction takes place across the ring. At the resonances where the 
conductance approaches the value $2$, the transmission probability $T$ 
goes to unity, since the relation $g=2T$ follows from the Landauer 
conductance formula (see Eq.~\ref{equ1} with $e=h=1$). All these 
resonant peaks are associated with the energy eigenvalues of the ring, 
and accordingly, we can say that the conductance spectrum manifests itself 
\begin{figure}[ht]
{\centering \resizebox*{8cm}{7cm}{\includegraphics{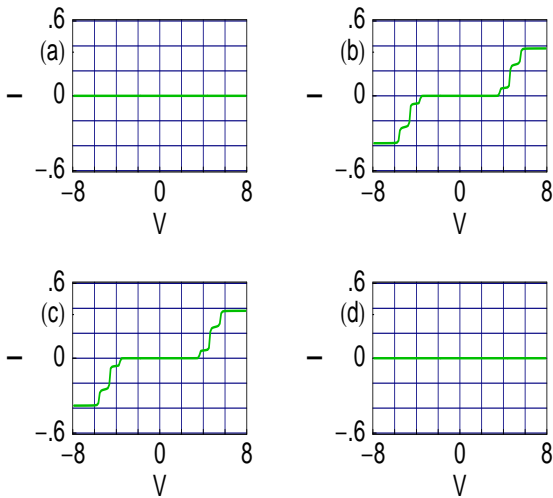}}\par}
\caption{(Color online). Current $I$ as a function of the bias voltage 
$V$ for a mesoscopic ring with $N=8$ and $\phi=0.5$ in the limit of 
weak-coupling. (a) $V_a=V_b=0$, (b) $V_a=2$ and $V_b=0$, (c) $V_a=0$ 
and $V_b=2$ and (d) $V_a=V_b=2$.}
\label{currlow}
\end{figure}
the electronic structure of the ring. Thus more resonant peaks can be
obtained for larger rings corresponding to their energy eigenvalues.
Now we justify the dependences of the gate voltages on the electron 
transport for these four different cases. The probability amplitude of
getting an electron across the ring depends on the quantum interference
of the electronic waves passing through the upper and lower arms of the 
ring. For the symmetrically connected ring i.e., when the two arms of the
ring are identical with each other, the probability amplitude is exactly
zero ($T=0$) for the flux $\phi=\phi_0/2$. This is due to the result of
the quantum interference among the two waves in the two arms of the ring, 
which can be obtained in a very simple mathematical calculation. Thus for 
the cases when both the two inputs
($V_a$ and $V_b$) are either low or high, the transmission probability
drops to zero. On the other hand, for the other two cases the symmetry
of the two arms of the ring is broken by applying the gate voltage either
in the atom $a$ or $b$, and therefore, the non-zero value of the 
transmission probability is achieved which reveals the electron conduction 
across the ring. Thus we can predict that the electron conduction takes
place across the ring if one, and only one, of the inputs to the gate is 
\begin{table}[ht]
\begin{center}
\caption{XOR gate behavior in the limit of weak-coupling. The current 
$I$ is computed at the bias voltage $6.02$.}
\label{table1}
~\\
\begin{tabular}{|c|c|c|}
\hline \hline
Input-I ($V_a$) & Input-II ($V_b$) & Current ($I$) \\ \hline 
$0$ & $0$ & $0$ \\ \hline
$2$ & $0$ & $0.378$ \\ \hline
$0$ & $2$ & $0.378$ \\ \hline
$2$ & $2$ & $0$ \\ \hline \hline
\end{tabular}
\end{center}
\end{table}
high, while if both the inputs are low or both are high the conduction is 
no longer possible. This feature clearly demonstrates the XOR behavior.
With these characteristics, we get additional one feature when the coupling 
strength of the ring to the electrodes increases from the low regime to
high one.
In the limit of strong ring-to-electrodes coupling, all these resonances 
get substantial widths compared to the weak-coupling limit. The results 
are shown in Fig.~\ref{condhigh}, where all the other parameters are 
identical to those in Fig.~\ref{condlow}. The contribution for
the broadening of the resonant peaks in this strong-coupling limit
appears from the imaginary parts of the self-energies $\Sigma_S$ and
$\Sigma_D$, respectively~\cite{datta}. Hence by tuning the coupling strength, 
we can get the electron transmission across the ring for the wider range of 
energies and it provides an important signature in the study of 
current-voltage ($I$-$V$) characteristics.

All these features of electron transfer become much more clearly visible
by studying the $I$-$V$ characteristics. The current passing through the 
ring is computed from the integration procedure of the transmission function 
$T$ as prescribed in Eq.~\ref{equ8}. The transmission function varies exactly 
similar to that of the conductance spectrum, differ only in magnitude by the 
factor $2$ since the relation $g=2T$ holds from the Landauer conductance 
formula Eq.~\ref{equ1}. As illustrative examples, in Fig.~\ref{currlow} we 
show the current-voltage characteristics for a mesoscopic ring with $N=8$ 
in the limit of weak-coupling.
For the cases when both the two inputs are identical with each other,
either low (Fig.~\ref{currlow}(a)) or high (Fig.~\ref{currlow}(d)), the 
current is zero for the entire bias voltages. This behavior is clearly 
understood from the conductance spectra, Figs.~\ref{condlow}(a) and (d), 
since the current is computed from the integration procedure of the 
transmission function $T$. For the two other cases where only one of the 
two inputs is high and other is low, a high output current is obtained
which are clearly described in Figs.~\ref{currlow}(b) and (c). From 
these figures it is observed that
the current exhibits staircase-like structure with fine steps as a 
function of the applied bias voltage. This is due to the existence of 
the sharp resonant peaks in the conductance spectrum in the weak-coupling
limit, since the current is computed by the integration method of the 
\begin{figure}[ht]
{\centering \resizebox*{8cm}{7cm}{\includegraphics{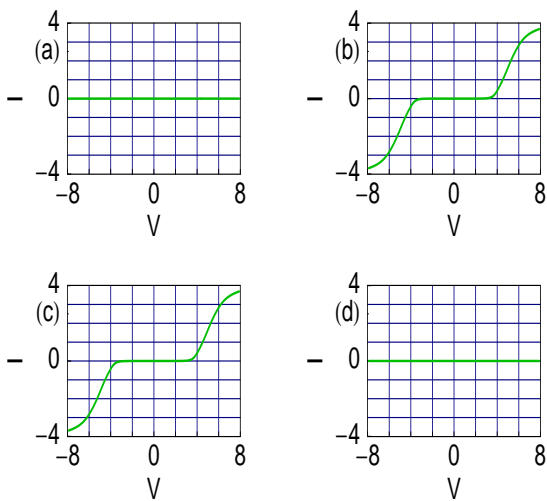}}\par}
\caption{(Color online). Current $I$ as a function of the bias voltage 
$V$ for a mesoscopic ring with $N=8$ and $\phi=0.5$ in the limit of 
strong-coupling. (a) $V_a=V_b=0$, (b) $V_a=2$ and $V_b=0$, (c) $V_a=0$ 
and $V_b=2$ and (d) $V_a=V_b=2$.}
\label{currhigh}
\end{figure}
transmission function $T$. With the increase of the bias voltage $V$, 
the electrochemical potentials on the electrodes are shifted gradually, 
and finally cross one of the quantized energy levels of the ring. 
Accordingly, a current channel is opened up which provides a jump
in the $I$-$V$ characteristic curve. Addition to these behaviors, it is
also important to note that the non-zero value of the current appears
beyond a finite value of $V$, so-called the threshold voltage ($V_{th}$).
This $V_{th}$ can be controlled by tuning the size ($N$) of the ring.
From these $I$-$V$ characteristics the behavior of the XOR gate response
is clearly visible. To make it much clear, in Table~\ref{table1}, we present 
a quantitative estimate of the typical current amplitude, computed at the 
bias voltage $V=6.02$, in this weak-coupling limit. It shows that $I=0.378$
only when any one of the two inputs is high and other is low, while for
the other cases when either $V_a=V_b=0$ or $V_a=V_b=2$, $I$ gets the value 
$0$. In the same analogy, as above, here we also discuss the $I$-$V$ 
characteristics for the strong-coupling limit. In this limit, the
current varies almost continuously with the applied bias voltage and
achieves much larger amplitude than the weak-coupling case as presented
in Fig.~\ref{currhigh}. The reason is that, in the limit of strong-coupling 
all the energy levels get broadened which provide larger current in the 
integration procedure of the transmission function $T$. Thus by tuning the 
strength of the ring-to-electrodes coupling, we can achieve very large 
\begin{table}[ht]
\begin{center}
\caption{XOR gate behavior in the limit of strong-coupling. The current 
$I$ is computed at the bias voltage $6.02$.}
\label{table2}
~\\
\begin{tabular}{|c|c|c|}
\hline \hline
Input-I ($V_a$) & Input-II ($V_b$) & Current ($I$) \\ \hline 
$0$ & $0$ & $0$ \\ \hline
$2$ & $0$ & $2.846$ \\ \hline
$0$ & $2$ & $2.846$ \\ \hline
$2$ & $2$ & $0$ \\ \hline \hline
\end{tabular}
\end{center}
\end{table}
current amplitude from the very low one for the same bias voltage $V$.
All the other properties i.e., the dependences of the gate voltages on the
$I$-$V$ characteristics are exactly similar to those as given in 
Fig.~\ref{currlow}. In this strong-coupling limit we also make a quantitative
study for the typical current amplitude, given in Table~\ref{table2}, where 
the current amplitude is determined at the same bias voltage ($V=6.02$) as
earlier. The response of the output current is exactly similar to that as
given in Table~\ref{table1}. Here the non-zero value of the current gets 
the value $2.846$ which is much larger compared to the weak-coupling case 
which shows the value $0.378$. From these results we can clearly manifest 
that a mesoscopic ring exhibits the XOR gate response.

\section{Concluding remarks}

To summarize, we have addressed XOR gate response in a mesoscopic metallic 
ring threaded by a magnetic flux $\phi$. The ring, composed of identical
quantum dots, is attached symmetrically to the source and drain. The upper 
and lower arms of the ring are subjected to the gate voltages $V_a$ and 
$V_b$, respectively those are taken as the two inputs of the XOR gate. A 
simple tight-binding model is used to describe the full system 
and all the calculations are done in the Green's function formalism. We 
have numerically computed the conductance-energy and current-voltage 
characteristics as functions of the ring-electrodes coupling strengths, 
magnetic flux and gate voltages. Very interestingly we have noticed that, 
for the half flux-quantum value of $\phi$ ($\phi=\phi_0/2$), a high 
output current ($1$) (in the logical sense) appears if one, and only one, 
of the inputs to the gate is high ($1$). On the other hand, if both the 
inputs are low ($0$) or both are high ($1$), a low output current ($0$) 
appears. It clearly manifests the XOR gate behavior and this aspect may be 
utilized in designing a tailor made electronic logic gate. In view of the
potential application of this XOR gate as a circuit element in an integrated 
circuit, we would like to point out that care should be taken during the
application of the magnetic field in the ring such that the other circuit 
elements of the integrated circuit are not effected by this field.

In this presentation, we have calculated all the results by ignoring 
the effects of the temperature, electron-electron correlation, disorder, 
etc. Due to these factors, any scattering process that appears in the 
arms of the ring would have influence on electronic phases, and, in 
consequences can disturb the quantum interference effects. Here we 
have assumed that, in our sample all these effects are too small, and 
accordingly, we have neglected all these effects in our present study.

The importance of this article is mainly concerned with (i) the simplicity 
of the geometry and (ii) the smallness of the size. To the best of our 
knowledge the XOR gate response in such a simple low-dimensional system 
has not been addressed earlier in the literature.

\end{document}